\documentclass[aps,pre,preprint,showpacs,groupedaddress,eqsecnum]{revtex4}%onecolumn

\begin{document}

\preprint{1}

\title{Modeling fine particle (dusty) plasmas and charge-stabilized colloidal suspensions
as inhomogeneous Yukawa systems}


\author{Hiroo Totsuji}
\email[]{totsuji-09@t.okadai.jp}
%\homepage[]{http://homepage3.nifty.com/totsuji/index2.html}
%\thanks{}
\affiliation{Graduate School of Natural Science and Technology, Okayama University, Okayama 700-8530, Japan}
%\affiliation{ISS Science Project Office, Institute of Space and Astonautical Science, Japan %Aerospace Exploration Agency (JAXA),
%Sengen 2-1-1, Tsukuba, Japan}
%\thanks{}

\date{\today}

\begin{abstract}
In order to give a basis for the structure and correlation analysis of fine particle (dusty) plasma and colloidal suspensions,
thermodynamic treatment of mixtures of macroscopic and microscopic charged particles
within the adiabatic response of the latter
is extended to include the case where
the system is finite and weakly inhomogeneous.
It is shown that
the effective potential for macroscopic particles is composed of two elements:
mutual Yukawa repulsion 
and 
a confining (attractive) Yukawa potential from their `shadow'
(the average charge density of macroscopic particles multiplied by the minus sign).
The result clarifies the relation between two approaches hitherto taken
where either a parabolic one-body potential is assumed
or
the average distribution is assumed to be flat with finite extension.
Since
the satisfaction of the charge neutrality is largely enhanced
by the existence of macroscopic particles,
the assumption of the flat electrostatic potential
and therefore flat average distribution of macroscopic particles 
in the domain of their existence
is expected to be closer to reality than the assumption of the parabolic potential
in that domain.

\end{abstract}

%\keywords{}

\maketitle


\section{introduction}%%%%%%%%%%%%%%%%%%%%%%%%%%%%%%%%%%%%%%%%%%%%%%%%%%%%%%

In many systems of physical interest,
there exist both macroscopic and microscopic particles
and
we usually apply
theoretical treatments based on the adiabatic response of microscopic particles.
Typical examples of such systems composed of charged particle are fine particle (dusty) plasmas and charge stabilized colloidal suspensions.
For instantaneous positions of fine particles (in the former) or colloidal particles (in the latter),
we take the statistical averages over the ambient plasma (electrons and ions) or positive and negative ions
to have the screening of the charges of macroscopic charged particles.
The effective interaction between them is then given
approximately by the Yukawa (Debye-H\"{u}ckel) or DLVO potential
which accompanies the factor $\exp(-r/\lambda)$ characterized by the screening length $\lambda$.

In this article,
we call macroscopic and microscopic particles simply `particles' and `background', respectively.
The adiabatic response of the background 
has been analyzed for the uniform system\cite{HF94,YR94}.
It is shown that,
in addition to the screening of Coulombic interactions between particles,
we have to take into account the charge neutrality of the system
which gives a confining potential due to the charge density of the background
canceling the average charge density of particles.
Let us consider the case of one species of particles.
If we consider only the effective interaction of Yukawa or DLVO type,
the system would have a tendency to expand (explode) 
with the pressure
which is a sum of the positive ideal gas pressure (of both particles and background) 
and 
the additional positive pressure coming from mutual Coulomb-like repulsion between particles.
When the charge neutrality of the system as a whole is properly taken into account,
however, the latter pressure is not positive definite:
For example, when particles are randomly distributed without correlation,
we have no average space charge
and
there should be no Coulombic contribution to the pressure.
Moreover,
from the result of weakly coupled plasmas described by the Debye-H\"{u}ckle theory, 
we expect that,
when the correlation develops between particles,
the Coulombic contribution to the pressure becomes {\it negative} 
and
the pressure is reduced from the ideal gas values.
In the case of strongly coupling,
the magnitude of the negative pressure even increases\cite{BST66}.

In numerical simulations
which have been useful in investigations of these systems\cite{RKG88,FH94,HFD97},
particles are usually regarded as interacting only via the Yukawa repulsion
without any consideration on the charge neutrality of the system.
The system of $N$ particles in a volume $V$
is regarded as a part of the infinite uniform system
with the number density $N/V$
and
the limit of $N, V \rightarrow \infty$ is taken within numerical possibilities.
When the average density $N/V$ is kept unchanged,
for example by the periodic boundary conditions with fixed volume,
correct distribution functions between particles can be obtained:
Since the change in $N/V$ is suppressed,
we are implicitly confining repelling particles
and
the effect of the charge neutrality need not to be explicitly reflected in simulations.
The consideration of the effect becomes necessary
only in the expression of the correlation energy
which is proportional to
\begin{equation}
\int d{\bf r} v(r) [g(r)-1].
\end{equation}
The integrand is the interaction potential $v(r)$ multiplied by the pair {\it correlation} function $g(r)-1$ of particles,
not by the pair {\it distribution} function $g(r)$.
Since the difference between them, 
$\int d{\bf r} v(r)$, 
is a finite (density-dependent) constant
for the screened potential $v(r)$,
the correction can be made separately,
even if the effect of the charge neutrality is not taken into account explicitly in the simulation.

The Helmholtz free energy for a given configuration of particles
can be calculated by taking the statistical average over microscopic particles.
The effective interaction is related to terms in the Helmholtz free energy
which includes the coordinates of particles.
As far as the uniform system is concerned,
other terms (not-including their coordinates) can be regarded as constants,
even if their values depend on number density and other parameters.
On the other hand,
when one considers the system of finite extension (or of finite geometrical size),
the latter terms also become important:
They are directly related to the confinement (the size and the shape) of the system.

There have been two different approaches
in treating finite systems.
One is to assume that the system is locally charge neutral
and
apply the result of infinite uniform system\cite{TTOT05a,TTOT05b,TT11}
and
the other is to assume some {\it ad hoc} confining potential\cite{HB06,KL08,GWL12}.
In the latter,
the parabolic potential of some sort has been usually adopted.

Both approaches have their own cases where they are applicable.
They are, however, not complete.
In the first approach,
the geometrical size of the system is to be determined by some {\it ad hoc} external origin,
for example,
by the radius of tube containing discharges or colloidal suspensions.
In the case of fine particle plasmas,
we also have stationary generation and transport of plasma (electrons and ions)
and
there exists finite space charge which produces the electric field
to maintain the ambipolar diffusion.
Though the space charge density is small relative to electron or ion density
(generally, proportional to the square of the ratio of the screening length to the system size),
the local charge-neutrality is not exactly satisfied.
In the second approach,
the geometrical size of the system is determined by the balance 
between the mutual repulsion of particles and the (often parabolic) confining potential.
When the system is charge-neutral,
the background neutralizing the system should have the effect 
which reinforces the confinement
and
this effect should also exist
even when the system is not exactly charge-neutral.
Neither of two approaches is thus complete
and,
in order to rectify such incompleteness,
it is necessary to construct a theory
which
includes both the non-uniformity of the system and the role of the background.
In this article,
we present some results in the case of weak non-uniformity.

We consider the system composed of particles (macroscopic charged particles) and background (microscopic charged particles).
In the case of fine particle plasmas,
the former is fine (dust) particles and the latter is ambient plasma of electrons and ions.
In the case of charge stabilized colloidal suspensions,
they are colloid particles and positive or negative ions, respectively.
Since the mass of macroscopic particles is much larger than those of microscopic ones,
we regard the latter as adiabatically responding to the instantaneous configuration of the former
and,
after the statistical average with respect to the background,
physical quantities become functions of the configuration of particles.

We assume that
both particles and background of our system 
are described by component-dependent temperatures
and
we can apply usual thermodynamic treatment to our system.
Since our system is often open in the sense that
we have mass and energy flows into and from our system,
especially in the case of fine particle plasmas,
the applicability of thermodynamics might be questionable.
There exist, however, no other well-established frameworks
and
we may justify our assumptions as a realistic approach.
In what follows,
we describe the system
in terms of fine particle plasmas.
One may easily interpret the results into the case of colloidal suspensions.

As fine particle plasmas,
we consider those generated by dc or rf discharge in inert gases such as Ar 
with the pressure $10\sim10^2$ Pa.
Typical densities of neutral gas atoms, plasma (electrons and ions), and fine particles
are of the order of $10^{15}\sim10^{16}$ ${\rm cm^{-3}}$, $10^8$ ${\rm cm^{-3}}$,
and
$10^5$ ${\rm cm^{-3}}$,
respectively.
As for neutral gas atoms,
their distribution is considered to be uniform and stationary without flows 
throughout the system.
 
We consider the case where
our system is in a stationary state
and
the distributions of electrons, ions, and particles are
characterized by the temperatures $T_e$, $T_i$, and $T_p$, respectively.
In most experiments,
$k_B T_e$ is of the order of a few eV,
while $T_i$ and $T_p$ are much lower than $T_e$
and
considered to be of the order of the room temperature.
(In colloidal suspensions,
both positive and negative ions are at the room temperature.
In what follows,
the use of the fact $T_e \gg T_i$ is explicitly mentioned
and
the conclusion common to both systems is independent of the fact.) 
We denote the (usually negative) charge on a particle by $-Qe (Q, e >0)$.

\section{local average and fluctuations}%%%%%%%%%%%%%%%%%%%%%%%%%%%%%%%%

In our system
there exist three different characteristic  lengths,
namely,
the size of the system $L$, 
the mean distance between particles $a_p$,
and 
the mean distances between electrons or ions $a_{e,i}$.
Their typical values are
$L \sim {\rm a\ few}\ {\rm cm}$,
$a_p \sim 10^{-2}\ {\rm cm}(=10^{2}\ {\rm \mu m})$,
and
$a_{e,i} \sim 10^{-3}\ {\rm cm}(=10\ {\rm \mu m})$
and
holds the inequality
\begin{equation}\label{basic-inequality}
L \gg a_p \gg a_{e,i}.
\end{equation}
Taking a distance $\ell$ so that
\begin{equation}
L \gg \ell \gg a_p \gg a_{e,i},
\end{equation}
we define the local average of a quantity $A({\bf r})$ at ${\bf r}$, $\overline{A}({\bf r})$,
by the space average over the domain of volume $\ell^3$ (with linear dimension of the order $\ell$)
centered at ${\bf r}$;
\begin{equation}
\overline{A}({\bf r}) 
\equiv 
{1 \over \ell^3} \int_{\ell^3\ {\rm centered\ at\ }{\bf r}} A ({\bf r}) d{\bf r}.
\end{equation}
The ${\bf r}$-dependence of $\overline{A}({\bf r})$ is characterized by $L$.
We also define the deviation (fluctuation) from the average $\delta A ({\bf r})$
by
\begin{equation}
\delta A ({\bf r}) \equiv A ({\bf r}) - \overline{A}({\bf r}).
\end{equation}

We denote the density of each component by $n_{e,\ i,\ \alpha}({\bf r})$.
The total charge density $\rho ({\bf r})$ is written as
\begin{equation}
\rho ({\bf r}) = \rho_p ({\bf r}) + \rho_{bg} ({\bf r}),
\end{equation}
where
$\rho_p ({\bf r})$ is the charge density of particles
\begin{equation}
\rho_p ({\bf r}) = (-Qe) n_p ({\bf r}) = (-Qe) \sum_{i=1}^N \delta({\bf r} - {\bf r}_i)
\end{equation}
and
$\rho_{bg} ({\bf r})$ the charge density of the background plasma
composed of electrons and ions
\begin{equation}
\rho_{bg} ({\bf r}) = (-e)n_e({\bf r})+en_i({\bf r}).
\end{equation}
The electrostatic potential $\Psi({\bf r})$ satisfies the Poisson's equation
\begin{equation}\label{Poisson-equation}
-\varepsilon_0 \Delta \Psi({\bf r})
=
\rho_p ({\bf r}) + \rho_{bg} ({\bf r})
=
e[-Q n_p ({\bf r}) - n_e({\bf r}) + n_i({\bf r})]
\end{equation}
and
we have
\begin{equation}
-\varepsilon_0 \Delta \overline{\Psi}({\bf r})=\overline{\rho}({\bf r})
=
\overline{\rho_p} ({\bf r}) + \overline{\rho_{bg}} ({\bf r})
\end{equation}
and
\begin{equation}\label{Poisson-delta}
-\varepsilon_0 \Delta \delta \Psi({\bf r})
=
\delta \rho_p ({\bf r}) + \delta \rho_{bg} ({\bf r})
\end{equation}
separately.
In (\ref{Poisson-delta}),
$\delta \rho_p({\bf r})$ includes the coordinates of particles
and
determines the polarization of electrons and ions,
$\delta n_e({\bf r})$ and $\delta n_i({\bf r})$, giving 
$\delta \rho_{bg} ({\bf r})= e[- \delta n_e({\bf r}) + \delta n_i({\bf r})]$.

As for the polarization of electrons and ions,
we adopt the approximation of the linear adiabatic response to the local potential fluctuation $\delta \Psi ({\bf r})$;
\begin{equation}\label{adiabatic-response-electron}
\delta n_e({\bf r}) \sim  \overline{n_e}({\bf r})[\exp(  {e \delta \Psi ({\bf r}) / k_B T_e})-1]
\sim \overline{n_e}({\bf r}) {e \delta\Psi ({\bf r}) \over k_B T_e},
\end{equation}
\begin{equation}\label{adiabatic-response-ion}
\delta n_i({\bf r}) \sim  \overline{n_i}({\bf r})[\exp( - {e \delta \Psi ({\bf r}) / k_B T_i})-1]
\sim - \overline{n_i}({\bf r})  {e \delta\Psi ({\bf r}) \over k_B T_i}.
\end{equation}
We then have
\begin{equation}\label{background}
\delta \rho_{bg} ({\bf r})
=  - \varepsilon_0 k_D^2({\bf r}) \delta \Psi ({\bf r})
\end{equation}
and
(\ref{Poisson-delta}) reduces to
\begin{equation}\label{Poisson-for-delta}
-\varepsilon_0 [\Delta-k_D^2({\bf r})] \delta\Psi({\bf r})
=
\delta \rho_p ({\bf r}).
\end{equation}
Here
$k_D({\bf r})$ is the (local) Debye wave number defined by
\begin{equation}
k_D^2({\bf r})
=
{e^2 \overline{n_e}({\bf r}) \over \varepsilon_0k_B T_e} + {e^2 \overline{n_i}({\bf r}) \over \varepsilon_0 k_B T_i}.
\end{equation}
Note that,
since we have generation/loss of the plasma and its transport by the (ambipolar) diffusion,
it is {\it  not} assumed that
$\overline{n_e}({\bf r}) \propto \exp(  {e \overline{\Psi}({\bf r}) / k_B T_e})$
nor
$\overline{n_i}({\bf r}) \propto \exp(  -{e \overline{\Psi}({\bf r}) / k_B T_i})$
for averages.
For typical electron and ion temperatures,
$1/k_D$ is much smaller than the system size $L$;
\begin{equation}
L \sim 1{\rm cm} 
\gg 
\left( {e^2 \overline{n_e}({\bf r}) \over \varepsilon_0k_B T_e} \right)^{-1/2} 
\sim 10^{-1}\ {\rm cm}
\gg
{1 \over k_D}
\sim \left({e^2 \overline{n_i}({\bf r}) \over \varepsilon_0 k_B T_i}\right)^{-1/2} \sim 10^{-2}\ {\rm cm}.
\end{equation}
We can thus regard the length $\ell$ satisfying the inequality 
\begin{equation}
L \gg \ell \gg \{1/k_D,\ a_p\} \gg a_{e,i}.
\end{equation}

When $k_D$ is ${\bf r}$-independent,
the solution for (\ref{Poisson-delta}) is given by
\begin{equation}
\delta\Psi({\bf r})
=
\int d{\bf r}'{\exp(-k_D |{\bf r}-{\bf r}'|) \over 4 \pi \varepsilon_0 |{\bf r}-{\bf r}'|}
\delta \rho_p ({\bf r}')
\end{equation}
and
$\delta\Psi({\bf r})$ is determined by the values of $\delta\rho_p({\bf r}')$
within distances of the order of $1/k_D$ from ${\bf r}$.
Since 
the position dependence of $1/k_D$ is characterized by $L$
and $L \gg 1/k_D$,
we may write the approximate solution for (\ref{Poisson-delta}) in the form (see Appendix A)
\begin{equation}\label{approximate-solution-Poisson}
\delta\Psi({\bf r})
\sim
\int d{\bf r}' u({\bf r},{\bf r}')\delta \rho_p ({\bf r}'),
\end{equation}
where
\begin{equation}
u({\bf r},{\bf r}') 
={\exp(-k_D^+ |{\bf r}-{\bf r}'|) \over 4 \pi \varepsilon_0 |{\bf r}-{\bf r}'|}
\end{equation}
and
\begin{equation}
k_D^+ = k_D [({\bf r}+{\bf r}')/2].
\end{equation}

\section{Effective interaction}%%%%%%%%%%%%%%%%%%%%%%%%%%%%%%%%%%%%%%%%%%%%%%%

\subsection{Helmholtz free energy for given configuration of particles}

Under the conditions of fixed volume and fixed temperatures of electrons and ions,
the work necessary to change the configuration of particles
is given by the change in the Helmholtz free energy of the system of electrons and ions\cite{LLSP80-1-20}.
The effective interaction energy for the system of fine particles is thus written as
\begin{equation}
U_{ex}
= F_{id}^{(e)} + F_{id}^{(i)} 
+ \left[{1 \over 2} \int d{\bf r} \rho ({\bf r}) \Psi ({\bf r}) - U_{s}\right].
\end{equation}
Here $F_{id}^{(e)}+F_{id}^{(i)}$ is the Helmholtz free energy of the background or the electron-ion plasma
\begin{equation}
F_{id}^{(e)}=k_B T_e \int d{\bf r} n_e ({\bf r}) \left(\ln [n_e ({\bf r}) \Lambda_e^3] -1 \right),
\end{equation}
\begin{equation}
F_{id}^{(i)}=k_B T_i \int d{\bf r} n_i ({\bf r}) \left(\ln [n_i ({\bf r}) \Lambda_i^3] -1 \right),
\end{equation}
$\Lambda_e$ and $\Lambda_i$ being the thermal de Broglie lengths.
In the third term,
we subtract the self-energy $U_s={(1 / 2)}\sum_{i=j}^N {(-Qe)^2 / 4 \pi \varepsilon_0 r_{ij}}$ included in the formal integral expression of the electrostatic energy.
We adopt the ideal gas value of the Helmholtz free energy
for electrons and ions:
Usually the coupling in the background plasma is very weak (the $\Gamma$ parameter is $10^{-3}$ to $10^{-4}$)
and the non-ideal effects are negligible,
while the coupling between fine particles can be very strong.
(Thermal de Broglie lengths are introduced only to define the unit volume of the phase space
and
have no relations to statistical properties of our classical system.)

We expand $F_{id}^{(e)}+F_{id}^{(i)}$ with respect to fluctuations.
Noting
$\int d{\bf r} \delta n_{e, i}  \ln [\overline{n_{e, i}} \Lambda_{e, i}^3],\ 
\int d{\bf r} \delta n_{e, i}  \sim 0$,
we have, to the second order,
\begin{equation}\label{Helmholts0}
F_{id}^{(e)}+F_{id}^{(i)} \sim
F_{id,0} +
{1 \over 2}\int d{\bf r}\left[k_B T_e 
{\delta n_e^2 ({\bf r}) \over \overline{n_e}({\bf r})} 
+  k_B T_i {\delta n_i^2 ({\bf r}) \over \overline{n_i}({\bf r})}\right]
=
F_{id,0} -
{1 \over 2} \int d{\bf r} \delta \rho_{bg}({\bf r}) \delta\Psi({\bf r}).
\end{equation}
Here
\begin{equation}
F_{id,0}=k_B T_e \int d{\bf r} 
\overline{n_e}({\bf r}) \left[\ln [\overline{n_e}({\bf r}) \Lambda_e^3] -1 \right]
+
k_B T_i \int d{\bf r} 
\overline{n_i}({\bf r}) \left[\ln [\overline{n_i}({\bf r}) \Lambda_i^3] -1 \right]
\end{equation}
and
(\ref{adiabatic-response-electron}), (\ref{adiabatic-response-ion}), and (\ref{background}) are used.
Since
$\int d{\bf r} \delta \rho_{p}  \overline{\Psi},\ 
\int d{\bf r} \delta \rho_{bg}  \overline{\Psi},\ 
\int d{\bf r} \overline{\rho_p}  \delta {\Psi},\ 
\int d{\bf r} \overline{\rho_{bg}}  \delta {\Psi}  \sim 0$,
the electrostatic energy is written as
\begin{equation}\label{electrostatic}
{1 \over 2} \int d{\bf r} \rho \Psi  -  U_s
={1 \over 2} \int d{\bf r} [\overline{\rho_p} + \overline{\rho_{bg}}]\overline{\Psi}
+{1 \over 2} \int d{\bf r} [ \delta \rho_p + \delta\rho_{bg}]\delta\Psi - U_s.
\end{equation}
From (\ref{Helmholts0}) and (\ref{electrostatic}), we have
\begin{equation}\label{effect-of-discreteness}
U_{ex}
=
F_{id,0} 
+ 
{1 \over 2} \int d{\bf r} [\overline{\rho_p}({\bf r})+\overline{\rho_{bg}}({\bf r})] \overline{\Psi}({\bf r})
+
\left[
{1 \over 2} \int d{\bf r} \delta\rho_p({\bf r}) \delta\Psi({\bf r})
- U_s
\right].
\end{equation}
By (\ref{approximate-solution-Poisson}),
the last term of (\ref{effect-of-discreteness}) is rewritten as
\begin{eqnarray}
& &{1 \over 2}\int\int d{\bf r} d{\bf r}' u({\bf r},{\bf r}')
[\rho_p({\bf r})-\overline{\rho_p}({\bf r})][\rho_p({\bf r}')-\overline{\rho_p}({\bf r}')] - U_s
\nonumber \\
&=&{(Qe)^2 \over 2}\sum_{i,j=1}^N u({\bf r}_i,{\bf r}_j) - U_s
-
(-Qe)\sum_{i=1}^N \int d{\bf r}' u({\bf r}_i,{\bf r}')
\overline{\rho_p}({\bf r}')
+
{1 \over 2}\int \int d{\bf r}d{\bf r}' u({\bf r},{\bf r}')\overline{\rho_p}({\bf r})\overline{\rho_p}({\bf r}').
\end{eqnarray}
First two terms on the right-hand side reduce 
to the mutual Yukawa repulsion and the free energy stored in the sheath;
\begin{equation}
{1 \over 2}\sum_{i,j=1}^N (Qe)^2u({\bf r}_i,{\bf r}_j)-U_s
=
{ (Qe)^2 \over 2}\sum_{i \neq j}^Nu({\bf r}_i,{\bf r}_j)
-{1 \over 2} \sum_{i=1}^N {(Qe)^2 k_D({\bf r}_i) \over 4 \pi \varepsilon_0}.
\end{equation}
The Helmholtz free energy is finally given by
\begin{eqnarray}\label{Helmholtz}
U_{ex}
&=&
F_{id,0} 
+ 
{1 \over 2} \int d{\bf r} [\overline{\rho_p}({\bf r})+\overline{\rho_{bg}}({\bf r})] \overline{\Psi}({\bf r}) \nonumber \\
& &+ 
\left[
{1 \over 2}\sum_{i \neq j}^N (Qe)^2 u({\bf r}_i,{\bf r}_j)
+
\sum_{i=1}^N (-Qe) \int d{\bf r}' u({\bf r}_i,{\bf r}')[-\overline{\rho_p}({\bf r}')]
\right] \nonumber \\
& &+
{1 \over 2}\int \int d{\bf r}d{\bf r}' 
u({\bf r},{\bf r}')\overline{\rho_p}({\bf r})\overline{\rho_p}({\bf r}') 
-{1 \over 2} \sum_{i=1}^N {(Qe)^2 k_D({\bf r}_i) \over 4 \pi \varepsilon_0}. 
\end{eqnarray}

\subsection{Potential for particles}

The averages, $\overline{\rho_p}({\bf r}), \overline{\rho_{bg}}({\bf r})$, and $\overline{\Psi}({\bf r})$, are to be determined
so as to be consistent with the plasma generation and loss and the ambipolar diffusion in the system.
Configuration-dependent terms in (\ref{Helmholtz}),
\begin{equation}\label{mutual interaction and confinement}
{1 \over 2}\sum_{i \neq j}^N {(Qe)^2 }u({\bf r}_i,{\bf r}_j)
+
\sum_{i=1}^N (-Qe) \int d{\bf r}' [- {\overline{\rho_p}({\bf r}')}]
u({\bf r}_i,{\bf r}')
- {1 \over 2} \sum_{i=1}^N {(Qe)^2 k_D({\bf r}_i) \over 4 \pi \varepsilon_0},
\end{equation}
describe the Helmholtz free energy for given distribution of particles $\{{\bf r}_i\}_{i=1,\dots N}$.
The integral in the second term
\begin{equation}
\int d{\bf r}'[- {\overline{\rho_p}({\bf r}')] }u({\bf r}_i,{\bf r}')
=
\int d{\bf r}'{[-\overline{\rho_p}({\bf r}')] \over 4 \pi \varepsilon_0 |{\bf r}_i-{\bf r}'|}
\exp(-k_D^+ |{\bf r}_i-{\bf r}'|)
\end{equation}
can be regarded as the Yukawa potential at ${\bf r}_i$ due to $[-\overline{\rho_p}({\bf r}')]$,
the (imaginary) charge density which exactly cancels the average particle charge density $\overline{\rho_p}({\bf r}')$:
We may call $[-\overline{\rho_p}({\bf r}')]$ the {\it ``shadow''} to $[\overline{\rho_p}({\bf r}')]$
emphasizing its difference from the background plasma
which really exists.
The charge density of the shadow has the sign opposite to particles
and
the potential due to the shadow is attractive for particles.
Particles are thus mutually interacting via the Yukawa repulsion
and,
at the same time,
confined by the attractive potential due to the shadow charge density $[-\overline{\rho_p}({\bf r}')]$\cite{TTOT05a,TTOT05b,TT11}.

\subsection{Infinite uniform system}%%%%%%%%%%%%%%%%%%%%%%%%%%%%

In the limit where $V,\ N \rightarrow \infty$ with $N/V$ kept constant,
we have
\begin{equation}
\delta\Psi({\bf r})
=
\int d{\bf r}'{\exp(-k_D |{\bf r}-{\bf r}'|) \over 4 \pi \varepsilon_0 |{\bf r}-{\bf r}'|}
\delta \rho_p ({\bf r}')
=
\sum_{i=1}^N {(-Qe) \over 4 \pi \varepsilon_0 |{\bf r}-{\bf r}_i|}\exp(-k_D |{\bf r}-{\bf r}_i|) 
- {\overline{\rho_p} \over \varepsilon_0 k_D^2}
\end{equation}
and
\begin{equation}\label{for-simulation}
{1 \over 2} \int d{\bf r} \delta \rho_p({\bf r}) \delta\Psi({\bf r}) - U_s
=
{1 \over 2} \sum_{i \neq j}^N v(r_{ij})
-{N \over 2} {(Qe)^2 (N/V) \over \varepsilon_0 k_D^2}
-{N \over 2} {(Qe)^2 k_D \over 4 \pi \varepsilon_0},
\end{equation}
where
\begin{equation}
v(r)={(Qe)^2 \over  4 \pi \varepsilon_0 r}\exp(-k_D r)
\end{equation}
and $k_D$ is position-independent.
In terms of the pair distribution function $g(|{\bf r}-{\bf r}'|)$,
we have
\begin{equation}\label{uniform-final-2}
U_{ex}
=
V n_ek_BT_e[\ln n_e\Lambda_e^3-1]+V n_ik_BT_i[\ln n_i\Lambda_i^3-1]
%F_{id,0} 
+
N \left[{n_p \over 2} \int d{\bf r} v(r) [g({\bf r})-1]
- {(Qe)^2 k_D \over 8 \pi \varepsilon_0}\right].
\end{equation}
The results thus reduce to the previous ones\cite{HF94,YR94}.

In numerical simulations
where
the average density of particles is kept unchanged,
the expression (\ref{for-simulation}) is used
and
only the first term is computed in each step.
Usually the number of particles is fixed
and
periodic boundary conditions are imposed.
Though distribution functions are correctly evaluated directly from such simulations,
we have to calculate the free energy based on the correlation energy (\ref{uniform-final-2}).

\subsection{Average with respect to particle distribution}%%%%%%%%%%%%%%%%%%%

Let us denote the statistical average with respect to the particles by $<\ \ >$.
Taking this average of (\ref{Helmholtz}) and noting that $<n_p({\bf r})>=\overline{n_p}({\bf r})$,
we have
\begin{eqnarray}\label{Helmholtz-average}
& &<U_{ex}>
=
F_{id,0} 
+ 
{1 \over 2} \int d{\bf r} [\overline{\rho_p}({\bf r})+\overline{\rho_{bg}}({\bf r})] \overline{\Psi}({\bf r}) \nonumber \\
&+& 
{1 \over 2}(Qe)^2 \int \int d{\bf r}d{\bf r}'  u({\bf r},{\bf r}') \left[
< \sum_{i \neq j}^N \delta({\bf r}-{\bf r}_i)\delta({\bf r}'-{\bf r}_j) >
- \overline{n_p}({\bf r})\overline{n_p}({\bf r}') 
\right] \nonumber \\ 
&-& <{1 \over 2} \sum_{i=1}^N {(Qe)^2 k_D({\bf r}_i) \over 4 \pi \varepsilon_0}>. 
\end{eqnarray}
The statistical average  
$< \sum_{i \neq j}^N \delta({\bf r}-{\bf r}_i)\delta({\bf r}'-{\bf r}_j) >$
is a function of both ${\bf r}$ and ${\bf r}'$.
Since
the average density changes with the scale length $L$
which is much larger than the mean distance between particles $a_p$,
this function depends mainly on ${\bf r}-{\bf r}'$
and the dependence on ${\bf r}$ or ${\bf r}'$ is weak.
We define a function $g[|{\bf r}-{\bf r}'|; ({\bf r}+{\bf r}')/2]$
with the arguments ${\bf r}-{\bf r}'$ and $({\bf r}+{\bf r}')/2$,
expecting the weak dependence on the latter: 
\begin{equation}
<\sum_{i \neq j}^N \delta({\bf r}-{\bf r}_i)\delta({\bf r}'-{\bf r}_j)>
\equiv
\overline{n_p}({\bf r})\ \overline{n_p}({\bf r}')
g[|{\bf r}-{\bf r}'|; ({\bf r}+{\bf r}')/2].
\end{equation}
In the uniform system,
$g[|{\bf r}-{\bf r}'|; ({\bf r}+{\bf r}')/2]$ reduces to the usual pair distribution function 
$g(|{\bf r}-{\bf r}'|).$
The third term on the right-hand side of  (\ref{Helmholtz-average}) is expressed by this function as
\begin{equation}
{1 \over 2}\int\int d{\bf r} d{\bf r}'
\overline{\rho_p}({\bf r})\ \overline{\rho_p}({\bf r}')[g(|{\bf r}-{\bf r}'|; ({\bf r}+{\bf r}')/2)-1]
u({\bf r},{\bf r}').
\end{equation}

The potential by the shadow (the second term of (\ref{mutual interaction and confinement})) influences the distribution
so that the average charge density approaches to $\overline{\rho_p}({\bf r})$:
The average force acting on the particles $i$ given by
\begin{equation}
(-Qe)^2<\sum_{j(\neq i)} [- \nabla_i u({\bf r}_i,{\bf r}_j)]  - \int d{\bf r}'[- \nabla_i u({\bf r}_i,{\bf r}')]{\overline{n_p}({\bf r}')}>
\end{equation}
reduces to zero
since the average particle distribution is $\overline{n_p}({\bf r})$.
(In principle,
there exists a possibility that,
when particles are strongly correlated,
the fluctuations have some effect on averages
and
averages and fluctuations need to be determined self-consistently.
In this article, however, we assume that
fluctuations are determined under given averages.)

\section{Discussions}%%%%%%%%%%%%%%%%%%%%%%%%%%%%%%%%%%%%%%%%%%%%%%%%%%%%%%

In this article,
the effective potential for particles is derived for given average distributions
which
are determined so as to be consistent with the generation/loss and the ambipolar diffusion of plasma.
The result generalizes the analysis of the infinite uniform system\cite{HF94,YR94} to finite weakly inhomogeneous systems.
Here
we note that,
in some cases,
the weakness of the inhomogeneity and finiteness of the system (or the distribution of particles)
are not compatible in a strict sense.
In those cases,
we have to be aware of the possibility that
the result may include errors
near the boundary.

We now discuss the relation to previous approaches to finite systems,
taking a typical example of  particles in the cylindrical positive column discharges.
In addition to facilitating simple geometry and symmetry,
we assume the generation of plasma in the bulk and the loss to the outer boundary (wall of apparatus) by ambipolar diffusion.
We expect
similar discussions apply also to more complicated cases.

\subsection{Case of negligible contribution to net charge density from particles}

When the contribution of particles to the net charge density is negligible,
the distribution of plasma and the electrostatic potential are determined 
independently of particles.
Then the plasma distribution and the potential are approximately expressed respectively by
\cite{FLJ66FFC74}
\begin{equation}
\overline{n_{e,i}}(R) \sim \overline{n_{e,i}}(R=0) J_0 (R/R_a),
\end{equation}
and
\begin{equation}\label{ambipolar potential}
\overline{\Psi}(R) \sim  {k_B T_e \over e} \ln  J_0 (R/R_a)
=  -{k_B T_e \over e} \left({R^2 \over 4 R^2_a} + \dots \right).
\end{equation}
Here
${\bf R}$ is the coordinates perpendicular to the symmetry axis,
$J_0$, the 0-th order Bessel function,
and $R_a$, a characteristic length of the order of (radial) system size
which is determined by the ambipolar diffusion coefficient and the rate of plasma generation.
The normalized charge density around $R=0$ is given by
\begin{equation}
0 < {-\varepsilon_0 \Delta \overline{\Psi}(R=0) \over e \overline{n_{e,i}}(R=0)}
\sim { T_e/T_i  \over R_a^2 (k_{De}^2 +  k_{Di}^2)} 
\sim { 1 \over R_a^2 k_{De}^2} 
\ll  1,
\end{equation}
where  $k_{Di}=[\overline{n_i(R=0)} e^2 /\varepsilon_0 k_B T_i]^{1/2}$ 
and $k_{De}=[\overline{n_e(R=0)} e^2 /\varepsilon_0 k_B T_e]^{1/2}$
are the ion and electron Debye wave numbers, respectively
(here we used the fact $T_e \gg T_i$ in usual cases).
Since usually $1/k_{De} \ll R_a$,
the quasi-charge-neutrality holds
and
we have slightly positive net charge density.
When the contribution of particles to the latter is negligible,
particles are considered to be also in this electrostatic potential
which is parabolic near the axis.
We thus have a model where
particles  in a parabolic confining potential mutually interact via the Yukawa repulsion,
corresponding to some of previous approaches to finite inhomogeneous system of particles\cite{HB06,KL08,GWL12}.

The distribution of particles is analyzed in Appendix B.
For the given value of the linear density of particles along the axis $n_{p, z}$,
\begin{equation}
n_{p, z}=\int d{\bf R} \overline{n_p}({\bf r}),
\end{equation}
particles are distributed within the radius $R_0$ such that
\begin{equation}\label{R_0}
{R_0 \over R_a} \sim 2\left ( {T_i \over T_e}{Qn_p \over n_e} \right)^{1/2}.
\end{equation}
Here
$n_p =\overline{n_p}(R=0)$, $n_e=\overline{n_e}(R=0)$, and $n_{p, z} \sim \pi n_p R_0^2$.
In order for the particle charges not to affect the potential,
$Qn_p/n_e \ll 1 /(k_{De}R_a)^2 \ll 1$
and
therefore we have $R_0/R_a \ll 1$
(we do not have the case where $T_i/T_e \gg 1$).

\subsection{Case of effective contribution to net charge density from particles}

With the increase of  $n_{p, z}$,
the radius $R_0$ and the charge density of particles $(-Qe)n_p$ increase
and,
when $(Qn_p)/n_e$ becomes not negligible compared with $1 /(k_{De}R_a)^2$,
we have to couple the particle charge density with the potential and therefore with electron and ion distributions.
Here,
it is important to note the point that
the potential and electron/ion distributions are related to the generation/loss and ambipolar diffusion of the plasma
and 
their characteristic scale length cannot become much smaller than the system size, in our case, $R_a$:
The diffusion flux is controlled by the electric field
and
is almost continuous
(we implicitly assume that the reconnection of plasma in the bulk is small).

If the$R-$dependence of the potential is characterized by $R_a$,
as given by (\ref{ambipolar potential}),
the particle distribution is limited to the radius $R_0$
which is still much smaller than $R_a$, as given by (\ref{R_0}).
On the other hand,
the potential structure reflecting the particle distribution
should have the characteristic scale of length of particle distribution $R_0$
which is much smaller than $R_a$.
This is a contradiction
which indicates that,
when we denote the characteristic scale length of the potential by $R'_a$,
we should have $R'_a \gg R_a$.

This means that
the electrostatic potential which reflects the existence of particles
becomes almost flat in the scale of $R_a$.
The fact that
the electrostatic potential becomes flatter 
in the domain where we have appreciable amount of particle charge
has been noticed\cite{SFA13}
and
its physical origin has been clarified\cite{HT14}:
The charge neutrality is controlled by the Debye wave number $k_D$
which includes the contribution from particles
\begin{equation}
k^2_{D}=k_{Di}^2 + k_{De}^2 + {(Qe)^2 n_p \over \varepsilon_0 k_B T_p} 
\sim  {(Qe)^2 n_p \over \varepsilon_0 k_B T_p}
\gg k_{Di}^2
\end{equation}
and,
since $Q \gg 1$,
the contribution to the Debye wave number is dominated by particles
even if $Q n_p/n_e$ is sufficiently small.
The normalized charge density is determined by $k_D$,
instead of $(k_{Di}^2 + k_{De}^2)^{1/2} \sim k_{Di}$,
and much reduced due to the very existence of particles
as
\begin{equation}
{T_e/T_i  \over R_a^2 k_{D}^2} \sim { T_e/T_i  \over R_a^2 k_{Dp}^2} \ll  { T_e/T_i  \over R_a^2 k_{Di}^2}
\sim { 1 \over R_a^2 k_{De}^2}.
\end{equation}
{\it The potential thus becomes almost flat (in the scale of the system size)
where particles exist.}

The simplest approximation in this case may be to assume that
the potential is completely flat
where particles exist.
In our previous analyses of structures and ordering of particles in finite systems\cite{TTOT05a,TTOT05b,TT11},
we have assumed the average particle distribution is uniform with finite extensions.
Noting the behavior of the potential under the existence of particles,
we may expect this treatment to be closer to reality than the assumption of the parabolic potential.

\section{Concluding Remark}%%%%%%%%%%%%%%%%%%%%%%%%%%%%%%%%%%%%%%%%%%%%%%%%%

The main result of our analysis is (\ref{Helmholtz}).
When the average distributions $\overline{\rho_p}({\bf r}),\ \overline{\rho_{bg}}({\bf r})$
and therefore the average electrostatic potential $\overline{\Psi}({\bf r})$
are determined so as to be consistent with the generation/loss and the ambipolar diffusion of plasma
of microscopic particles,
structures of macroscopic particles are formed by the terms dependent on the configuration of macroscopic particles, (\ref{mutual interaction and confinement}).
Since the existence of macroscopic particles with large charges
is expected to enhance the satisfaction of the the charge neutrality 
to a much better accuracy\cite{SFA13,HT14}
and give much flatter electrostatic potential
than the case without them,
the uniform distribution of  $\overline{\rho_p}$ and therefore the shadow  $-\overline{\rho_p}$
up to some extension\cite{TTOT05a,TTOT05b,TT11} may be justified as a first approximation.

\acknowledgements

The author wishes to thank members of Working Group of ISAS/JAXA
(Institute of Space and Astronautic Sciences, Japan Aerospace Exploration Agency) 
especially Drs. K. Takahashi and S. Adachi for useful discussions.
He also thanks the ISS Science Project Office of ISAS/JAXA,
where this work started,
for kind supports.

%\bibliography{totsuji.bib}

\section*{Appendix A}%%%%%%%%%%%%%%%%%%%%%%%%%%%%%%%%%%%%%%%%%%

For the equation to be solved
\begin{equation}\label{equation}
[\Delta-k_D^2({\bf r})] \delta\Psi({\bf r})
=
-{\delta \rho_p ({\bf r}) \over \varepsilon_0},
\end{equation}
we first take the kernel $u^{(1)}({\bf r}, {\bf r}')$ defined by
\[
u^{(1)}({\bf r}, {\bf r}') \equiv {\exp(-k_D({\bf r}') |{\bf r}-{\bf r}'|) \over 4 \pi  |{\bf r}-{\bf r}'|}
\]
and
consider an approximate solution
\begin{equation}\label{solution}
 \int d{\bf r}' u^{(1)}({\bf r}, {\bf r}'){\delta \rho ({\bf r}') \over \varepsilon_0}.
\end{equation}
Since
\[
[ \Delta - k_D^2({\bf r})] u^{(1)}({\bf r}, {\bf r}')
 =
{\exp(-k_D({\bf r}) |{\bf r}-{\bf r}'|) \over 4 \pi}
\Delta {1 \over  |{\bf r}-{\bf r}'|}
+ [ k_D^2({\bf r}')- k_D^2({\bf r})] u^{(1)}({\bf r}, {\bf r}'),
\]
we have
\[
[ \Delta - k_D^2({\bf r})] \int d{\bf r}' u^{(1)}({\bf r}, {\bf r}'){\delta \rho ({\bf r}') \over \varepsilon_0}
\]
\[
 =
-{\delta \rho_p ({\bf r}) \over \varepsilon_0}
+ \int d{\bf r}'  [ k_D^2({\bf r}')- k_D^2({\bf r})] u^{(1)}({\bf r}, {\bf r}'){\delta \rho ({\bf r}') \over \varepsilon_0}.
\]
Noting that 
the characteristic scale of length for $k^2_D$ (or the density) is $L$
and the effective range of the kernel is of the order of $1/k_D$,
the relative error is estimated to be of the order of $1/k_D L \ll 1$
or
\begin{equation}\label{solution1}
[ \Delta - k_D^2({\bf r})] \int d{\bf r}' u^{(1)}({\bf r}, {\bf r}'){\delta \rho ({\bf r}') \over \varepsilon_0}
\sim - \left(1 + {\cal O}{1 \over k_D L}\right) {\delta \rho_p ({\bf r}) \over \varepsilon_0}.
\end{equation}
When we take the kernel
\[
u^{(2)}({\bf r}, {\bf r}') \equiv {\exp(-k_D({\bf r}) |{\bf r}-{\bf r}'|) \over 4 \pi  |{\bf r}-{\bf r}'|},
\]
we have
\[
[ \Delta - k_D^2({\bf r})] u^{(2)}({\bf r}, {\bf r}')
\]
\[
 =
{\exp(-k_D({\bf r}) |{\bf r}-{\bf r}'|) \over 4 \pi} \left[
\Delta {1 \over  |{\bf r}-{\bf r}'|}
+  {{\bf r}-{\bf r}' \over  |{\bf r}-{\bf r}'|} \cdot \nabla k_D^2({\bf r})
-\Delta  k_D({\bf r}) +  |{\bf r}-{\bf r}'|  (\nabla k_D({\bf r}))^2
\right],
\]
\[
[ \Delta - k_D^2({\bf r})] \int d{\bf r}' u^{(2)}({\bf r}, {\bf r}'){\delta \rho ({\bf r}') \over \varepsilon_0}
\]
\[
 =
-{\delta \rho_p ({\bf r}) \over \varepsilon_0}
+ \int d{\bf r}' u^{(2)}({\bf r}, {\bf r}') ({\bf r}-{\bf r}' ) \cdot \nabla k_D^2({\bf r})
{\delta \rho ({\bf r}') \over \varepsilon_0}
\]
\[
+  \int d{\bf r}' u^{(2)}({\bf r}, {\bf r}')
[ -|{\bf r}-{\bf r}'| \Delta  k_D({\bf r}) +  |{\bf r}-{\bf r}'|^2  (\nabla k_D({\bf r}))^2]
{\delta \rho ({\bf r}') \over \varepsilon_0},
\]
and similarly
\begin{equation}\label{solution2}
[ \Delta - k_D^2({\bf r})] \int d{\bf r}' u^{(2)}({\bf r}, {\bf r}'){\delta \rho ({\bf r}') \over \varepsilon_0}
\sim - \left(1 + {\cal O}{1 \over k_D L}\right) {\delta \rho_p ({\bf r}) \over \varepsilon_0}.
\end{equation}
The value of $k_D$ in the exponential function can thus be either taken at ${\bf r}$ or ${\bf r}'$
and therefore at  $({\bf r}+{\bf r}')/2$;
In fact,
$k_D({\bf r}) \sim k_D({\bf r}') \sim k_D [({\bf r}+{\bf r}')/2] 
\sim [k_D ({\bf r}) +k_D({\bf r}')]/2 $
when
$|{\bf r}-{\bf r}'| < 1/k_D$.

We note that,
in order to have (\ref{solution1}) or (\ref{solution2}) at any point ${\bf r}$ in the system,
it is important to have $k_D({\bf r})$ or $k_D({\bf r}')$ in the argument of the exponential function of the kernel.
When we fix the value of $k_D$ at some point ${\bf r}_0$
and
adopt $\exp(-k_D({\bf r}_0) |{\bf r}-{\bf r}'|) / 4 \pi  |{\bf r}-{\bf r}'|$ 
instead of $u^{(1)}({\bf r}, {\bf r}')$ or $u^{(2)}({\bf r}, {\bf r}')$,
\[
{k_D^2({\bf r}) -k_D^2({\bf r}_0) \over k_D^2({\bf r}_0)}
\]
can be of the order of unity 
and,
even if we have $k_D L \gg 1$,
(\ref{equation}) is satisfied only around  ${\bf r}={\bf r}_0$.

\section*{Appendix B}

We take the $z$-axis along the axis
and
express the coordinates in real space by $({\bf R}, z)$.
The distribution of particles may be approximately estimated by assuming the uniform distribution
with the radius $R_0$.
The mutual interaction energy (per length along $z$) of uniformly distributed Yukawa particles
is given by
\[
\pi {(-Qe)^2 n_p^2 \over \varepsilon_0}{R^2_0 \over k^2_D}
\left[{1 \over 2} - K_1(k_DR_0)I_1(k_DR_0) \right],
\]
where $K_1$ and $I_1$ are the modified Bessel functions.
The energy due to the potential (\ref{ambipolar potential}) is given by
\[
{\pi \over 8} (Qe ) n_p {k_B T_e \over e} {R^4_0 \over R_a^2}.
\]
When $ k_DR_0 \gg 1$,
\[
\left[{1 \over 2} - K_1(k_DR_0) I_1(k_DR_0) \right] \sim {1 \over 2}
\]
and,
for given values of $n_{p,z}= n_p \pi R_0^2$, 
the total energy is minimum when
\[
\left({R_0 \over R_a}\right)^2 
= {4 \over  \pi \varepsilon_0}{Qe^2n_{p,z} \over k_B T_e}{1 \over k^2_D R^2_0}
\]
or
\[
\left({R_0 \over R_a}\right)^2 = {4 \over  \varepsilon_0}{Qe^2 n_p \over k_B T_e}{1 \over k^2_D}
= 4 {T_i \over T_e}{Qn_p \over n_{e,i}}.
\]

\end{document}